\documentclass{aa}  

\usepackage{url}
\usepackage{graphicx}
\usepackage{txfonts}
\usepackage{lscape}
\usepackage{amsmath}
\usepackage{multirow}
\bibpunct{(}{)}{;}{a}{}{,} % to follow the A&A style (Copied this in)

\usepackage{placeins}

\begin{document} 

   \title{A ${\sim} 6$ Mpc overdensity at $z \simeq 2.7$ detected along a pair of quasar sight lines: filament or protocluster?\thanks{Based on observations with X-shooter on the Very Large Telescope at the European Southern Observatory under program 089.A-0855.}}  % (UT2-Kueyen)
   %\subtitle{}

   \author{Hayley Finley \inst{1}
          \and
          Patrick Petitjean \inst{1}
          \and 
          Pasquier Noterdaeme \inst{1} 
          \and 
          Isabelle P\^{a}ris \inst{2}
          }

   \institute{Institut d'Astrophysique de Paris, CNRS-UPMC, UMR7095, 
              98bis bd Arago, 75014 Paris, France -- \email{finley@iap.fr}
         \and
         	 INAF - Osservatorio Astronomico di Trieste, Via G. B. Tiepolo 11, I-34131 \\
             }

   \date{\today}

%\abstract{}{}{}{}{} 
% 5 {} token are mandatory 

\abstract{
Simulations predict that gas in the intergalactic medium (IGM) is distributed in filamentary structures that connect dense galaxy clusters and form the cosmic web. 
These structures of predominantly ionized hydrogen are difficult to observe directly due to their lack of emitting regions. 
We serendipitously detected an overdensity of log~$N$(\ion{H}{I})~$> 18.0$ absorbers at $z \simeq 2.69$ along the lines of sight toward a pair of background quasars. Three main absorption regions spanning ${\sim}2\,000\ \rm km\ s^{-1}$ (corresponding to $6.4\ {h_{70}}^{-1}$~Mpc proper) are coincident in the two lines of sight, which are separated by ${\sim} 90\ {h_{70}}^{-1}$~kpc transverse proper distance.  
Two regions have [Fe/H] $< -1.9$ and correspond to mild overdensities in the IGM gas.
The third region is a sub-DLA with [Fe/H]~$=-1.1$ that is probably associated with a galaxy.
We discuss the possibility that the lines of sight probe along the length of a filament or intercept a galaxy protocluster.
}


   \titlerunning{A ${\sim} 6$ Mpc overdensity at $z \simeq 2.7$ detected along a pair of quasar sight lines}
   \authorrunning{H. Finley et al.}
   
   \maketitle

%
%________________________________________________________________

\section{Introduction}

Filamentary structures that emerge both from the large-scale distribution of galaxies and in cosmological simulations are iconic for the cosmic web \citep{1996Natur.380..603B}. Filaments and sheets outline vast, extremely underdense regions known as voids, before meeting at nodes that coincide with matter-rich galaxy clusters.
% The filamentary structures that are iconic of the cosmic web \citep{1996Natur.380..603B} outline vast extremely underdense regions (voids) before meeting at nodes that coincide with matter-rich galaxy clusters. 
Various techniques are used to identify structures in cosmological simulations and trace filaments \citep[e.g.,][]{2010MNRAS.409..156B, 2011MNRAS.413.2288M, 2011MNRAS.414..384S, 2012MNRAS.422...25S, 2014MNRAS.441.2923C}, the longest of which span more than $100\ h^{-1}$~Mpc. Segments connecting two clusters are relatively straight with typical lengths of $5 - 20\ h^{-1}$~Mpc and radial profiles that fall off beyond $2\ h^{-1}$~Mpc \citep{2005MNRAS.359..272C, 2010MNRAS.407.1449G, 2010MNRAS.408.2163A}.

At low redshift, filament finding techniques applied to the Sloan Digital Sky Survey \citep[SDSS;][]{2000AJ....120.1579Y} galaxy distribution measure maximum lengths comparable to simulations: $60 - 110\ h^{-1}$ Mpc \citep{2011MNRAS.411..332P, 2014MNRAS.438.3465T}. The majority of galaxies lie within $0.5\ h^{-1}$~Mpc of the filament axis \citep{2014MNRAS.438.3465T}.
Another strategy is to look for evidence of filamentary structures that connect a particular galaxy cluster to the cosmic web. Observations of clusters at $z \sim 0.5$ reveal that they are embedded in filaments extending more than $14\ h^{-1}$~Mpc \citep{2007MNRAS.379.1546T, 2012MNRAS.421.1949V}. % 20 Mpc
Complimentary to using galaxies as tracers, filaments can also be directly detected from weak gravitational lensing signals \citep{2010MNRAS.401.2257M}. \cite{2012MNRAS.426.3369J} unambiguously identify a filament with projected length ${\sim} 3.3\ h^{-1}$~Mpc (3D length $13.3\ h^{-1}$~Mpc) feeding into a massive galaxy cluster at $z = 0.55$.

%V4{
%At high redshift, absorption line spectroscopy probes diffuse \ion{H}{I} in the intergalactic medium (IGM), which traces large-scale structure. IGM clouds imprint numerous successive \ion{H}{I} absorptions in the spectra of high-redshift quasars, creating the so-called Lyman-alpha (Ly$\alpha$) forest. % that can be used to determine properties of the intervening gas. 
%\cite{2013A&A...552A..96B} and \cite{2013JCAP...04..026S} identify the baryon acoustic oscillation peak in the Ly$\alpha$ forest of $\langle z \rangle = 2.3$ quasars observed for the Sloan Digital Sky Survey III \citep[SDSS-III;][]{2011AJ....142...72E} Baryon Oscillation Spectroscopic Survey \citep[BOSS;][]{2013AJ....145...10D} by measuring correlations in the transmitted flux. The position of this peak at $100 - 130\ h^{-1}$~Mpc corresponds to a characteristic length in the large-scale matter distribution. 

%Quasar lines of sight with small angular separations reveal correlations on the scale of $< 5\ h^{-1}$~Mpc that likely arise from filaments \citep{2006MNRAS.370.1804C}. }

At high redshift, diffuse \ion{H}{I} in the intergalactic medium (IGM) imprints absorptions in the spectra of background quasars and creates the Lyman-alpha (Ly$\alpha$) forest. 
Correlations on scales $< 5\ h^{-1}$~Mpc comoving in the Ly$\alpha$ forests of quasar lines of sight (LOS) with small angular separations \citep[e.g.,][]{1998A&A...339..678D, 2003MNRAS.341.1279R, 2006MNRAS.370.1804C, 2006MNRAS.372.1333D, 2008MNRAS.385..519S, 2010MNRAS.407.1290C} likely arise from filaments.
Reconstruction methods applied to simulated and observed IGM absorptions recover the topology of this low-density gas at  $z \sim 2$ \citep{2008MNRAS.386..211C, 2014MNRAS.440.2599C}. 
However, little is known observationally about the topology of the IGM, and the actual \ion{H}{I} gas distribution may be less filamentary than simulated structures \citep{2012ApJ...750...67R}. % than in simulations what simulations suggest
Currently, the source density limits our ability to resolve cosmic web filaments. 
\cite{2014ApJ...788...49L} suggest that observing programs with existing $8 - 10$~m telescopes could achieve the source density necessary to obtain a resolution of ${\sim}3 - 4\ h^{-1}$~Mpc over cosmologically interesting volumes. However, the next generation of 30~m-class telescopes will best address the challenge of resolving filaments \citep{2009astro2010S.286S, 2013arXiv1310.3163M, 2014arXiv1406.6369E}

It is clear from these studies that quasar LOS intersect structures in the cosmic web. 
While they most often pass through the filament width, certain LOS foreseeably probe along the length.
Here we present \ion{H}{I} absorptions indicative of the gaseous environment within a filament. We detect multiple, consecutive absorptions at $z \simeq 2.69$ with $\log\ N$(\ion{H}{I})~(cm$^{-2}$)~$> 18.0$ that span nearly $2\,000\ \rm km\ s^{-1}$  and are coincident in \textit{both} LOS toward a pair of quasars separated by about 11\arcsec. % 90$~kpc proper distance .

% (16.4~$h^{-1}$~Mpc comoving)
% ($0.23\ h^{-1}$~Mpc comoving)

We describe the quasar spectra in Section 2, including how the close LOS pair was identified, and analyze the absorptions in each LOS in Section 3. In Section 4, we discuss evidence for whether the LOS intercept a galaxy protocluster or probe along the length of a filament. We use a $\Lambda$CDM cosmology with $\Omega_{\Lambda} = 0.73$, $\Omega_{\text{m}} = 0.27$, and $\rm H_{0} = 70\ km\ s^{-1}\ Mpc^{-1}$ \citep{2011ApJS..192...18K}. %In all cited values, $h = 0.7$.

%
%__________________________________________________________________

\section{Data}

The targeted quasars relevant to this work are SDSS J091338.30-010708.7 at $z \sim 2.75$ ($r = 20.49$) and J091338.96-010704.6 at $z \sim 2.92$ ($r = 20.38$). We refer to them as the foreground (FG) and background (BG) quasar accordingly. Their angular separation is 10.74\arcsec, which corresponds to $87.8\ {h_{70}}^{-1}$~kpc proper distance ($0.32\ {h_{70}}^{-1}$~Mpc comoving) at $z = 2.69$. These quasars were identified in the publicly available data release 9 quasar catalog \citep[DR9Q;][]{2012A&A...548A..66P} from the SDSS-III \citep{2011AJ....142...72E} Baryon Oscillation Spectroscopic Survey \citep[BOSS;][]{2013AJ....145...10D}. 

Initial interest in the pair was due to a damped Ly$\alpha$ absorption (DLA) in the BG LOS at the redshift of FG quasar, which offers an opportunity to study the host galaxy environment in absorption \citep[][and in preparation]{2013A&A...558A.111F}. %, (Finley et al., in preparation). 
In the low-resolution BOSS spectrum, an additional DLA with $\log N(\ion{H}{I}) = 21.05$ at $z_{\rm abs} = 2.680$ is flagged in the BG LOS \citep{2012A&A...547L...1N}. A corresponding system appears in the FG BOSS LOS, but the low column density excludes it from the catalog of $\log N(\ion{H}{I}) \geq 20$ absorbers. Motivated by the absorption systems, we pursued a higher resolution analysis of these LOS.

The quasars were observed in service mode in spring 2013 with X-shooter on the 8.2m Kueyen (UT2) telescope at the European Southern Observatory as part of a program (ESO 089.A-0855, P.I. Finley) targeting non-binary quasar pairs with small angular separations. The X-shooter spectrograph has UVB, VIS, and NIR arms that allow simultaneous observations across the full wavelength range from 300~nm to 2.5~$\mu$m. 
%We observed in long-slit stare mode (11\arcsec\ slit) with slit widths of 1.0\arcsec\ for the UVB and 0.9\arcsec\ for the VIS and NIR arms. 
%The quasars were observed in service mode over a total of four nights in 2013: March 30 - April 1 (FG), April 11-12 (BG), and April 30 - May 1 (BG). 
The total exposure times were $2 \times 3000$~s (1.67~h) for the FG quasar and $5 \times 3720$~s (5.17~h) for the BG quasar. % We dedicated two Observing Blocks (OBs), each with 3000~s exposure times in the UVB and VIS, to the FG quasar and five OBs, each with 3720~s exposure time in the UVB and VIS, to the BG qusar. The NIR observations were conducted in stare mode with five (six) 600~s exposures per OB for the FG (BG) quasar.

The data were reduced with version 2.2.0 of the ESO X-shooter pipeline \citep{2010SPIE.7737E..28M}. 
%\textit{The reduction process will be fully described in Finley et al.\ (in prep).} 
The bias level for the raw UVB and VIS frames was corrected by calculating the bias from the overscan region. %, as is the default pipeline behavior. 
Cosmic rays in the science exposures were flagged with the \cite{2001PASP..113.1420V} Laplacian edge detection method. 
After background subtraction, the science exposures were divided by the master flat for the appropriate arm, created from flat frames taken during the same day of observations.
%A master flat for each arm was created from flat frames taken the same day as the observations and applied to the science exposures. % within a day of the science exposures
Sky emission lines were then subtracted using the technique from \cite{2003PASP..115..688K}. 
Each spectral order was rectified from image space to wavelength space, using the 2D wavelength solution obtained from calibration frames. 
The individual 2D orders were extracted and merged, with pixel values weighted by the inverse variance of the corresponding errors in the overlapping regions.
1D spectra were obtained via standard extraction in the pipeline. 
%  however the implementation of their removal in the pipeline was unsatisfactory.

The extracted 1D spectra from different exposures were shifted to the vacuum-heliocentric reference frame and combined with an inverse variance weighted average. As in \cite{2012A&A...540A..63N}, we correct a 0.2\ \AA\ shift between the UVB and VIS spectra. The signal-to-noise ratio is 47 (21) at 5350 \AA\ and 38 (15) at 8100 \AA\ in the BG (FG) spectrum. We find that the resolution in the VIS spectra ($\rm R \approx 11\,000$), measured from the width of telluric absorption lines, is higher than the nominal resolution ($\rm R = 8800$), since the seeing was smaller than the 0.9\arcsec\ slit width. The resolution in the UVB (1.0\arcsec\ slit width) is likewise approximately $\rm R \approx 6400$.

%
%______________________________________________________________

\begin{table*}%[!bhtp]
\caption{Column Densities [log($N$/$\rm cm^{-2}$)] for components of the log~$N$(\ion{H}{I})~$= 20.2\ \rm cm^{-2}$ sub-DLA detected in the BG quasar LOS. Velocities (km~s$^{-1}$) are relative to the BG-C system redshift, $z = 2.6894$. The main components of both \ion{O}{I} and \ion{C}{II} are saturated.}
\label{table:NHI_C}
\centering
\begin{tabular}{ c  r  l  c  c  c  l  c  c  c  c  c  c  c }
\hline \hline
$z$ & $v$ & \multicolumn{1}{c}{\ion{O}{I}} & $\sigma_{\ion{O}{I}}$  & \ion{Si}{II} & $\sigma_{\ion{Si}{II}}$ & \multicolumn{1}{c}{\ion{C}{II}} & $\sigma_{\ion{C}{II}}$ & \ion{Al}{II} & $\sigma_{\ion{Al}{II}}$ & \ion{Al}{III} & $\sigma_{\ion{Al}{III}}$ & \ion{Fe}{II} & $\sigma_{\ion{Fe}{II}}$\\
\hline
2.687165 & -180 & 14.52: & 0.04 & 13.31 & 0.08 & 14.75: & 0.11 & 12.22 & 0.06 & 11.69 & 0.35 & 12.95 & 0.05 \\
2.688148 & -100 & 15.09: & 0.02 & 14.17 & 0.02 & 14.92: & 0.03 & 12.90 & 0.03 & 11.93 & 0.29 & 13.73 & 0.01 \\
2.688912 &  -38 & 16.23: & 0.33 & 14.11 & 0.04 & 15.10: & 0.44 & 13.35 & 0.10 & 12.21 & 0.13 & 14.16 & 0.03 \\
2.689329 &   -4 & 17.59: & 0.44 & 14.72 & 0.07 & 14.61: & 0.66 & 12.95 & 0.06 & 12.09 & 0.17 & 13.74 & 0.03 \\
2.689953 &   47 & 14.89: & 0.03 & 14.04 & 0.02 & 15.07: & 0.03 & 12.76 & 0.03 & 11.57 & 0.90 & 13.58 & 0.02 \\
2.692830 &  280 & 14.16 & 0.03 & 13.44 & 0.06 & 14.19 & 0.02 & 12.23 & 0.05 &    -- &   -- & 13.06 & 0.04 \\
\hline
\end{tabular}
\end{table*}

\begin{table}
\caption{\ion{H}{I} column densities [log($N$/$\rm cm^{-2}$)] for components along the BG and FG quasar LOS. Velocities (km~s$^{-1}$) are relative to the BG-C system redshift, $z = 2.6894$.}
\label{table:BGFG_NHI}
\centering
\begin{tabular}{c  r  c  c  } 
\hline \hline 
%\multirow{2}{*}{$v$, test} & BG & FG \\
Comp. & \multicolumn{1}{c}{$v$} & BG & FG \\
Name  & (km $\rm s^{-1}$) & log $N$(\ion{H}{I}) & log $N$(\ion{H}{I}) \\
\hline
BG-A1 & -1681 & 19.90 &    -- \\  %BG-A1
      & -1675 &    -- & 14.32 \\  %FG-A1
FG-A1 & -1550 &    -- & 18.46 \\  %FG-A2
BG-A2, FG-A2 & -1432 & 19.67 & 18.57 \\  %BG-A2, FG-A3
      & -1176 & 14.82 & 14.66 \\  %BG-HI, FG-HI 
      &  -893 &    -- & 15.30 \\  %FG-B1
BG-B  &  -834 & 18.41 &    -- \\  %BG-B1 
FG-B  &  -800 &    -- & 18.77 \\  %FG-B2
      &  -719 & 15.20 & 14.54 \\  %BG-B2, FG-B3
      &  -680 &    -- & 14.39 \\  %FG-B4
      &  -451 & 14.92 &    -- \\  %BG-HI
      &  -444 &    -- & 14.50 \\  %FG-HI
      &  -351 &    -- & 14.15 \\  %FG-HI 
      &  -190 &    -- & 14.75 \\  %FG-C1
      &   -78 &    -- & 15.90 \\  %FG-C2
BG-C  &     0 & 20.15 &    -- \\  %BG-C
      &    26 &    -- & 15.75 \\  %FG-C2
      &   294 & 16.22 &    -- \\  %BG-HI
\hline
\end{tabular}
\end{table}

\section{Absorption Systems}

\begin{figure*}[!p]
	\centering
	\includegraphics[width=\textwidth]{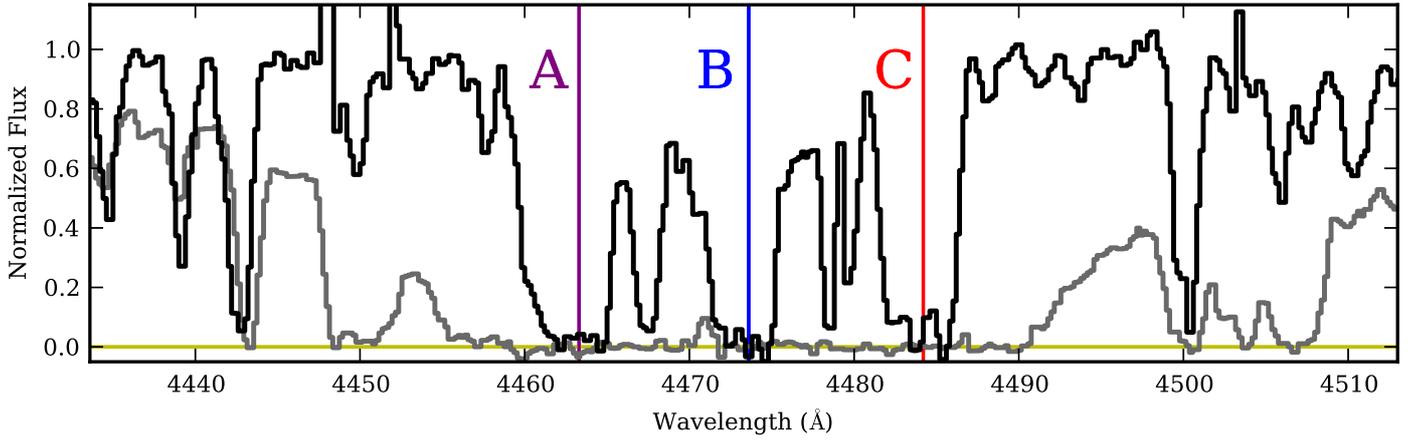} % width=18cm		                             
	\caption{%Velocity plot with the FG spectrum (black) overplotted on the BG spectrum (gray) relative to $z = 2.68938$. Significant \ion{H}{I} absorptions are coincident in the two spectra between $+300$ and $-2\,000\ \rm km\ s^{-1}$. 
	Portion of the Ly$\alpha$ forest with significant, coincident \ion{H}{I} absorptions along both LOS. The FG spectrum (black) is overplotted on the BG spectrum (gray). The main absorption regions are labelled A (purple), B (blue), and C (red).}
    \label{fig:oplot}
\end{figure*}

\begin{figure*}[!p]
	\centering
	\includegraphics[width=\columnwidth]{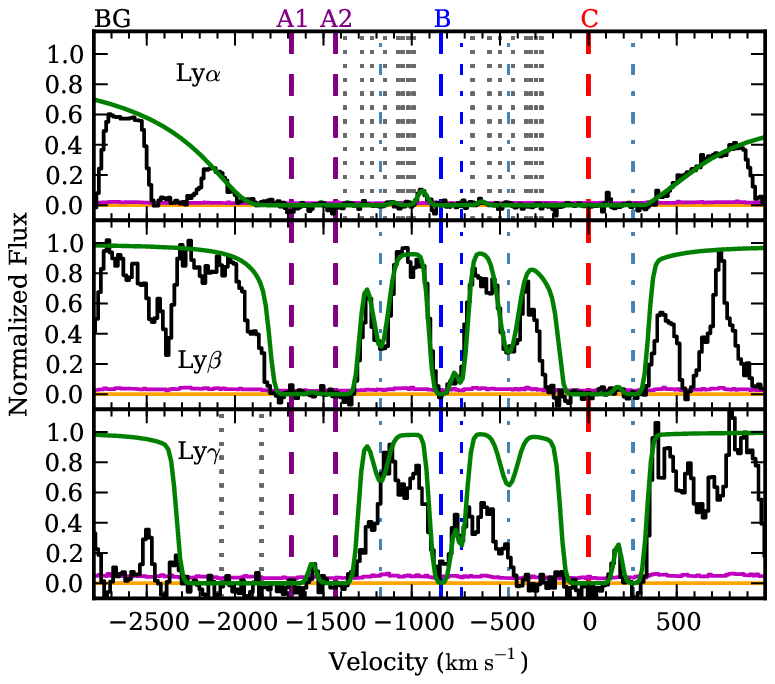}\hspace{2 mm} % width=18cm
	\includegraphics[width=\columnwidth]{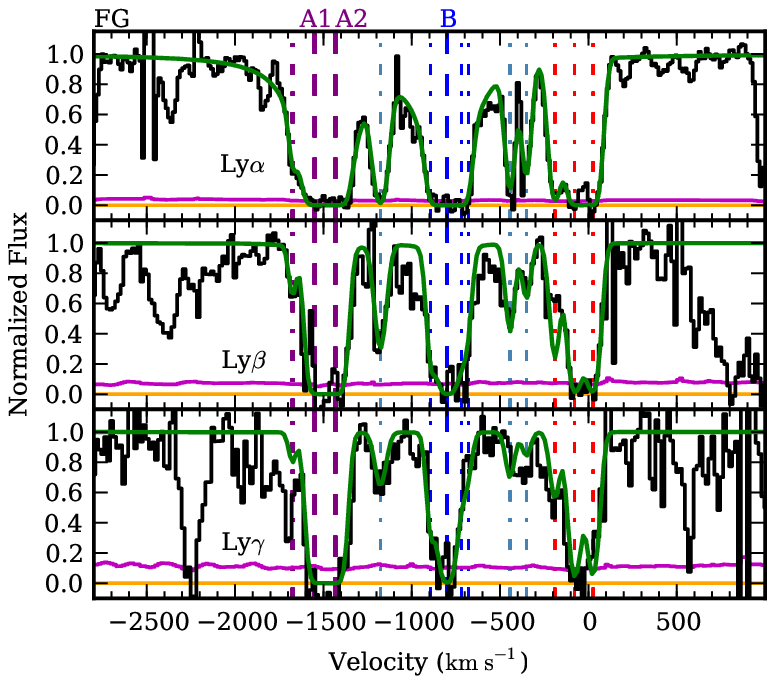}
	\caption{Fits to Ly$\alpha$ (top), Ly$\beta$ (middle), and Ly$\gamma$ (bottom) \ion{H}{I} absorptions in the BG (left) and FG (right) spectra. Dashed purple, blue, and red lines mark the log~$N$(\ion{H}{I})~$> 18.0$ components in regions A, B, and C, while dash-dotted purple, blue, and red lines indicate the weaker components within the respective regions. Dash-dotted blue-gray lines signal low column density components between the three main regions that are also part of the absorption structure. Dotted gray lines in the BG-Ly$\alpha$ panel indicate blended components from \ion{Si}{II}~$\lambda\lambda$1190, 1193 absorptions associated with a $z \simeq 2.75$ DLA.}
	% Components in regions A (dashed purple), B (dashed blue), and C (dashed red) correspond to absorptions in the BG spectrum. Low column density components (dash-dotted blue-gray) between the three main regions are also part of the absorption structure.
    \label{fig:HI_Fits}
\end{figure*}

\begin{figure*}[!p]
	\centering
	\includegraphics[width=\textwidth]{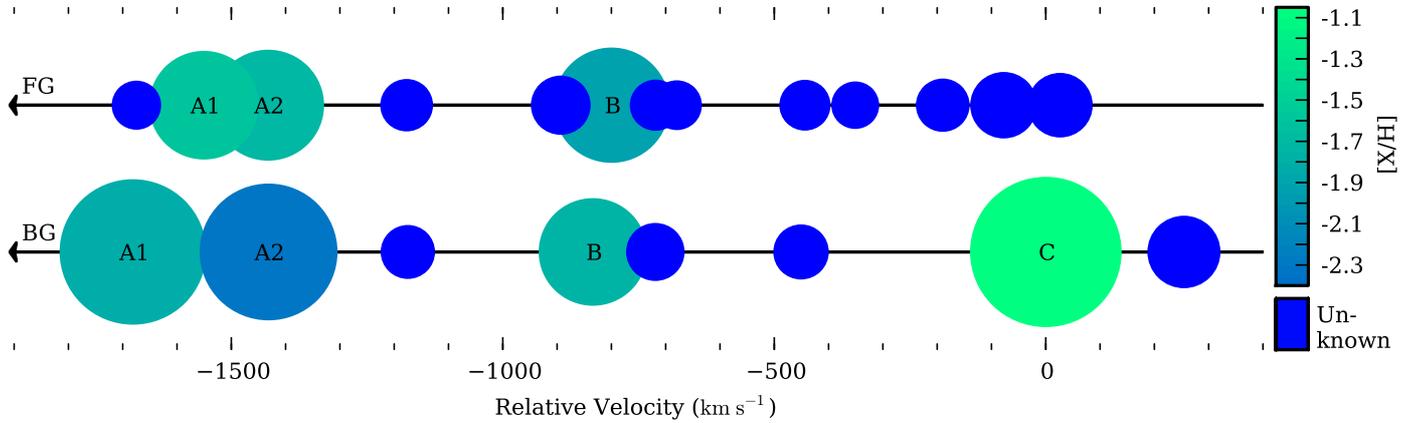} 
	\caption{Diagram of \ion{H}{I} clouds distributed along the FG and BG quasar LOS. Circle sizes scale with the \ion{H}{I} column density such that twice the area represents eight times as much $N$(\ion{H}{I}) (Area~$= N(\ion{H}{I})^{1/3}$). Low column density clouds, for which no metallicity is measured, are dark blue, while brighter, greener colors indicate clouds with higher metallicities. Zero velocity is at $z = 2.6894$.  }
		 % signify
		 % Dark blue circles indicate low column density absorptions. Brighter, greener circles indicate clouds with higher metallicity.
    \label{fig:diagram}
\end{figure*}

We identify consecutive intervening \ion{H}{I} absorptions spanning $\Delta v \simeq 2\,000\ \rm km\ s^{-1}$ at $z \simeq 2.69$ that are coincident in both LOS toward the J0913-0107 non-binary quasar pair. A proper distance of ${\sim} 90\ {h_{70}}^{-1}$ kpc at this redshift separates the FG and BG quasar LOS. Three main absorption regions are denoted A, B, and C in the two spectra (Figure \ref{fig:oplot}). We fit the entire absorption structure with the VPFIT package\footnote{\url{http://www.ast.cam.ac.uk/~rfc/vpfit.html}} to obtain system redshifts and column densities for the components (Figure~\ref{fig:HI_Fits}). Seven absorptions have $\log\ N$(\ion{H}{I}) (cm$^{-2}$) $> 18.0$, and we refer to them by the LOS, absorption region, and component number: BG-A1, BG-A2, BG-B, BG-C, FG-A1, FG-A2, and FG-B. We discuss the absorption systems in each LOS.

\subsection{Background Quasar Line of Sight}

\begin{figure}
	\centering
	\includegraphics[width=\columnwidth]{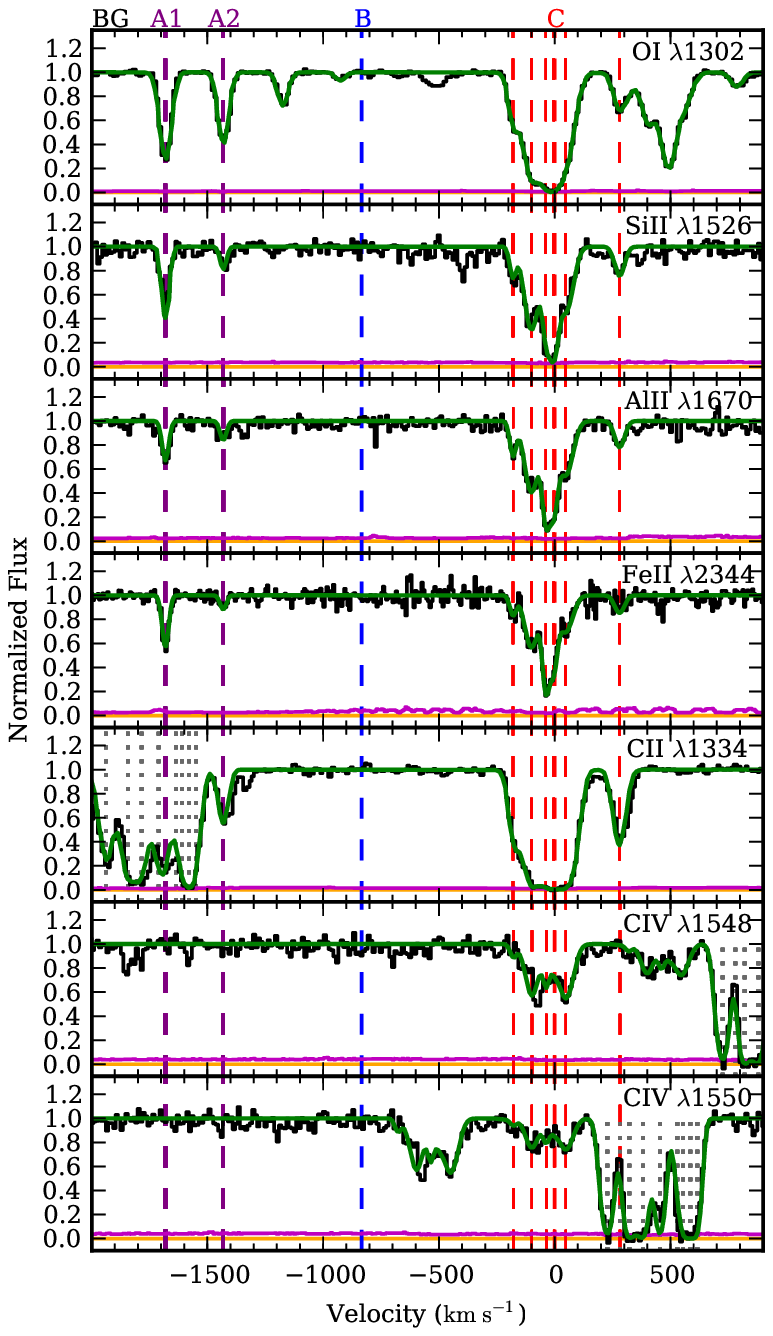}
	\caption{Fits to metal absorption lines in the BG quasar spectrum. Dashed purple, blue, and red lines mark components in regions A, B, and C. Thin dashed red lines indicate the six individual low-ionization components associated with region C. \ion{C}{IV} absorptions are not detected for the BG-A1 and BG-A2 components, and no metal absorption lines associated with the BG-B component are detected. \ion{Si}{II}~$\lambda$1304 absorptions appear directly to the right of the \ion{O}{I}~$\lambda$1302 absorptions in the uppermost panel. The BG-A1 \ion{C}{II} component is blended with the \ion{Si}{II}~$\lambda$1304 absorption from a $z \simeq 2.75$ DLA (dotted gray lines), but the BG-A2 component is unaffected. Dotted gray lines in the \ion{C}{IV} $\lambda\lambda$1548, 1550 panels likewise indicate components from the \ion{Si}{II}~$\lambda$1526 absorption associated with the same $z \simeq 2.75$ DLA. Zero velocity is at $z = 2.6894$, and the 1-$\sigma$ error on the flux is shown in magenta.}
	% Fits to the \ion{Si}{II}~$\lambda$1304 absorptions are also shown in the uppermost panel, directly to the right of the \ion{O}{I}~$\lambda$1302 absorptions.
    \label{fig:BGmet_Fits}
\end{figure}

\subsubsection{\ion{H}{I} Absorption Systems}

The \ion{H}{I} absorption profiles for the components in regions A, B, and C are constrained from fitting Ly$\alpha$ -- Ly$\delta$ in the UVB spectrum (Figure \ref{fig:HI_Fits}, left). The redshifts for components BG-A1 ($z = 2.6688$), BG-A2 ($z = 2.6718$), and BG-C ($z = 2.6894$) are fixed based on the fits to their associated low-ionization metal transitions (Figure~\ref{fig:BGmet_Fits}). The absorption in region C is a $\log\ N$(\ion{H}{I}) $= 20.2 \pm 0.1$ sub-DLA, and the associated metals are fitted with six components. The main absorption, with five components, spans ${\sim}225$~km~s$^{-1}$, and the sixth component at 280~km~s$^{-1}$ is redshifted relative to the $z = 2.6894$ system redshift. The BG-C redshift is the average of the six components weighted by their \ion{Fe}{II} column densities, which are given in Table~\ref{table:NHI_C}. 

The flux in the vicinity of Ly$\alpha$ is almost completely absorbed, except for a small peak separating region A from regions B and C at $-950\ \rm km\ s^{-1}$. Components BG-A1 and BG-A2 are both sub-DLAs, with $\log\ N$(\ion{H}{I}) $= 19.9 \pm 0.1\ \rm and\ 19.7 \pm 0.3$ respectively. Strong \ion{Si}{II}~$\lambda\lambda$1190, 1193 absorptions from a $z \simeq 2.75$ DLA blend with the \ion{H}{I} absorptions and contribute to the extended zero-level flux. The components in region B are more apparent in the Ly$\beta$ profile, and when they are included the fit to Ly$\alpha$ recovers the small peak near $-950\ \rm km\ s^{-1}$. The strong component labelled BG-B (Figure \ref{fig:HI_Fits}, left) is a $\log\ N$(\ion{H}{I})~$= 18.4 \pm 0.2$ Lyman limit system (LLS). All eight \ion{H}{I} absorptions fitted in the BG spectrum are listed with their velocity offsets relative to the BG-C component in Table~\ref{table:BGFG_NHI}.

\subsubsection{Abundances} 
\label{subsec:AbundBG}

Table \ref{table:Met_BG} gives abundances for the LLS and sub-DLA systems in the three regions. The abundances are calculated with respect to solar values \citep{2003ApJ...591.1220L} following the convention [X/H] $\equiv \log$(X/H) -- $\log$(X/H)$_{\odot}$. % [X/H]$_{\odot}$

The BG-A1 and BG-A2 metal absorptions are single-component, and we detect \ion{O}{I}, which is a good indicator of the metallicity. Charge transfer processes imply that \ion{O}{I} and \ion{H}{I} are tightly related \citep{1971ApJ...166...59F}. 
Since both the \ion{O}{I}~$\lambda$1302 and \ion{O}{I}~$\lambda$1039 transitions are detected for the BG-A1 component, the absorption line fit is well-constrained.
The oxygen abundance is [O/H]~$ = -1.19 \pm 0.34$. The Si, Al, and Fe abundances are slightly lower with [X/H]~$ = -1.7$, -2.1, and -1.9 respectively. The BG-A1 \ion{C}{II}~$\lambda$1334 absorption is blended with \ion{Si}{II}~$\lambda$1304 from the $z = 2.75$ DLA. We estimate the de-blended C abundance, [C/H]~$ = -1.83 \pm 0.59$, by fixing the DLA $N$(\ion{Si}{II}) from other transitions and imposing the same FWHM as for the other BG-A1 absorptions.
% Fixing the DLA $N$(\ion{Si}{II}) from other transitions, an estimate of the de-blended C abundance is [C/H]~$ = -1.83 \pm 0.59$.

The \ion{O}{I}~$\lambda$1039 transition is blended for the BG-A2 component, but the absorptions are non-saturated. 
The oxygen abundance, [O/H]~$ = -1.56 \pm 0.43$, is again slightly higher than the Si, C, Al, and Fe abundances [X/H]~$ \lesssim -2.1$. The BG-A2 \ion{C}{II}~$\lambda$1334 absorption is redder than the DLA \ion{Si}{II}~$\lambda$1304 absorption and unaffected by blending.
% We use \ion{O}{I}~$\lambda$1302 in both systems, along with \ion{O}{I}~$\lambda$1039 in BG-A1 (the BG-A2 line is blended) to estimate O metallicities. Errors are large because of saturation.
%                                                                         indicate that the region A absorptions

The [C/O] values, $-0.64 \pm 0.68$ for BG-A1 and $-0.54 \pm 0.55$ for BG-A2, follow the trend where, in low-metallicity systems, [C/O] increases as [O/H] decreases \citep{2011MNRAS.417.1534C, 2014MNRAS.440..307D}. 

No metal transitions corresponding to the \ion{H}{I} absorptions in region B are detected (Figure~\ref{fig:BGmet_Fits}) to a limit of $\log\ N$(\ion{O}{I}) $\leq 13.0 \pm 0.1$. To obtain this estimate, we use the average FWHM from the detected BG-A1 and BG-A2 \ion{O}{I} components and limit the absorption strength according to the noise in the flux. The upper limit on the [O/H] abundance is $-1.80 \pm 0.25$.

The abundances for the region C sub-DLA, [Si/H]~$ = -0.71 \pm 0.11$, [Al/H]~$ = -0.89 \pm 0.08$, and [Fe/H]~$ = -1.13 \pm 0.18$, are somewhat higher than the average value for intervening DLAs at $z \simeq 2.69$, $\langle Z \rangle = -1.24 \pm 0.12$ \citep{2012ApJ...755...89R}.
% For the region C sub-DLA, the abundance [Si/H]~$ = -0.71 \pm 0.11$ is somewhat higher than the mean metallicity, $\langle Z \rangle = -1.24 \pm 0.12$, of intervening DLAs at this redshift \citep{2012ApJ...755...89R}. 
The enhanced [Si/Fe] value, $= 0.41 \pm 0.21$, is typical of intervening DLAs at this redshift and is likely due to dust depletion \citep{2002ApJ...566...68P, 2002A&A...391..407V}.
Absorptions BG-A1, BG-A2, and BG-B all have abundances approximately an order of magnitude lower than that of BG-C.

%
%______________________________________________________________

\subsection{Foreground Quasar Line of Sight}

\begin{figure}
	\centering
	\includegraphics[width=\columnwidth]{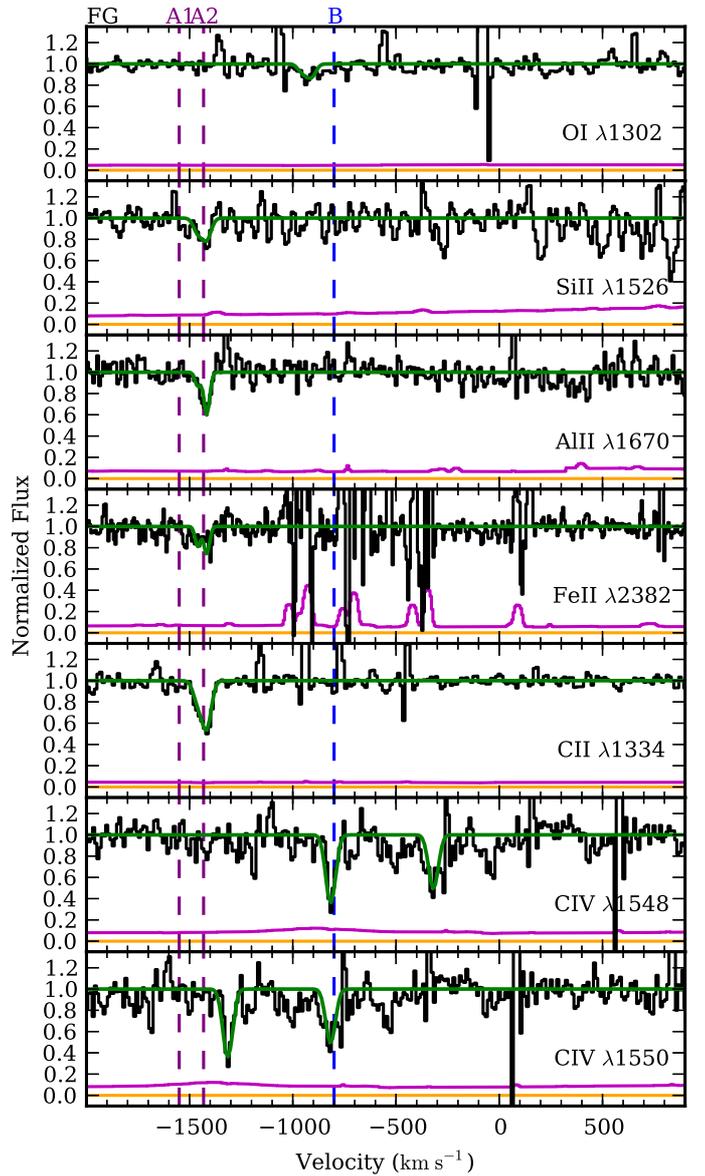}
	\caption{Fits to metal absorption lines in the FG quasar spectrum. Dashed purple and blue lines mark the strong \ion{H}{I} components in regions A and B. Weak low-ionization transitions (\ion{C}{II}, \ion{Si}{II}, \ion{Al}{II}, \ion{Fe}{II}) associated with FG-A2 are fitted with two components. No \ion{C}{IV} is detected in region A to a limit of log~$N$(\ion{C}{IV})~$ < 13.2 \pm 0.1$. For FG-B, only \ion{C}{IV} is detected. Zero velocity is at $z = 2.6894$, and the 1-$\sigma$ error on the flux is shown in magenta.}
	% Fits to the \ion{Si}{II}~$\lambda$1304 absorptions are also shown in the uppermost panel, directly to the right of the \ion{O}{I}~$\lambda$1302 absorptions.
    \label{fig:FGmet_Fits}
\end{figure}

\subsubsection{\ion{H}{I} Absorption Systems}

\ion{H}{I} absorptions in the FG quasar Ly$\alpha$ forest have a similar structure as the systems in the same redshift range in the BG quasar spectrum (Figure \ref{fig:oplot}). Weaker components separate the main concentrations of \ion{H}{I} in regions A, B, and C. 
We fit thirteen components to the Ly$\alpha$ -- Ly$\epsilon$ transitions for this absorption structure (Figure \ref{fig:HI_Fits}, right). 
Their column densities are listed in Table~\ref{table:BGFG_NHI}, along with the velocity offset relative to the BG-C component redshift.
 Three components, FG-A1, FG-A2, and FG-B, are in the LLS range, with $\log\ N$(\ion{H}{I}) $\rm = 18.5,\ 18.6,\ and\ 18.8$ respectively, all with $\sigma_{N(\ion{H}{I})} \simeq 0.2$. The remaining ten components are all below $\log\ N$(\ion{H}{I}) $= 16.0$. 
The highest column density components, FG-A1, FG-A2, and FG-B, are aligned with strong absorptions in regions A and B of the BG quasar LOS (Figure \ref{fig:diagram}). The FG-A1 component is between the BG-A1 and BG-A2 components, whereas the FG-A2 is exactly aligned with BG-A2 and FG-B is offset from BG-B by less than 35~km~s$^{-1}$.                                 % along the FG LOS
In region C, three lower column density components with $\log\ N$(\ion{H}{I}) $\simeq 14.8$, 15.9, and 15.8 occur within 200~km~s$^{-1}$ of the BG-C sub-DLA.

\subsubsection{Abundances}

% The only low-ionization metals detected are possible weak \ion{C}{II} and \ion{Si}{II}~$\lambda$1260 absorptions associated with the FG-A3 \ion{H}{I} component (Figure~\ref{fig:FGmet_Fits}). 
%The absorptions have asymmetric profiles stemming from two components. 
%However, \ion{Si}{II}~$\lambda$~1260 is blended with an unrelated, higher-redshift absorption.
%The upper limit on the C abundance is [C/H]~$\leq -0.86 \pm 0.42$.

Low-ionization metals are detected only for the FG-A2 \ion{H}{I} component (Figure~\ref{fig:FGmet_Fits}). The \ion{C}{II}, \ion{Si}{II}, \ion{Al}{II}, and \ion{Fe}{II} absorptions are fitted with two components, as required to follow the \ion{C}{II} profile. The absorptions are weak, however, and often difficult to distinguish from the noise. Upper limits on the abundances are [C/H]~$\leq -0.70 \pm 0.48$, [Si/H]~$\leq -0.48 \pm 0.40$, [Al/H]~$\leq -0.50 \pm 0.45$, and [Fe/H]~$\leq -1.08 \pm 0.47$. 
Since the FG-A2 \ion{H}{I} column density is log~$N$(\ion{H}{I})~$= 18.6$, the gas is not predominantly neutral and ionization corrections are likely significant. Both \ion{C}{II} and \ion{Si}{II} can be associated with the ionized gas.  

To obtain a reliable metallicity indicator, we estimate an upper limit of $\log\ N$(\ion{O}{I}) $\leq 13.5 \pm 0.1$ for the three LLS, FG-A1, FG-A2, and FG-B, using the same process as in Section~\ref{subsec:AbundBG}. Their corresponding metallicity limits are [O/H]~$\leq -1.7$, $-1.8$, and $-2.0$.

%
%______________________________________________________________

\section{Discussion and Conclusions}

%\cite{2013ApJ...775...81R} demonstrate that close LOS pairs can be used to measure the Jeans scale, which determines the clumpiness of the IGM, constrain the temperature and ionization history of the IGM. simulated spectra of close LOS pairs demonstrate a technique using simulated spectra of close LOS pairs to measure the Jeans scale, which determines the clumpiness of the IGM, and thereby constrain the temperature and ionization history of the IGM. directly measures the Jeans scale from coherence of correlated Ly$\alpha$ forest absorption demonstrate a technique for directly measuring the Jeans scale based on characterizing the coherence of correlated Ly$\alpha$ forest absorptions in simulated close LOS pairs 

% Efforts to identify close quasar pairs are motivated by 
% identified employed

% corresponding
We studied coincident \ion{H}{I} absorptions that occur in LOS toward the FG and BG quasars in the J0913-0107 pair.
Samples of close quasar pairs have been employed to measure quasar clustering when the redshift differences are negligible \citep[e.g.,][]{2010ApJ...719.1672H} and to investigate quasar host galaxy environments when the redshifts are offset \citep[e.g.,][]{2013ApJ...776..136P}. Despite growing statistics, very few strong coincident absorptions along the LOS have been reported. % (studied in detail?) % for LOS with small angular separations
% LOS pairs with strong coincident absorptions have been studied in detail. 
\cite{2007MNRAS.378..801E} analyzed a binary quasar pair featuring coincident DLA/sub-DLA systems at both $z = 2.66$ and $z = 2.94$ along the two LOS. 
The transverse distance between the LOS, ${\sim} 110\ h^{-1}_{70}$~kpc, is very similar to that of the LOS presented here; however, the separation between the absorbers along the LOS is an order of magnitude larger than the extent of the coincident absorption region in the J0913-0107 pair. 
The absorbers also have high [Zn/H] abundances. %, $-0.71 \pm 0.22$, $-0.51 \pm 0.33$, and $<0.51$, with no metallicity measured for the weakest log~$N$(\ion{H}{I})~$= 19.7$ sub-DLA. 
After comparing with cosmological simulations, the authors determined that the coincident absorptions are more likely due to groups of two or more galaxies than individual large galaxies.

In this work, the velocity separation, metallicities, and kinematics for coincident \ion{H}{I} absorptions along the studied region in the two J0913-0107 quasar spectra suggest that their LOS probe the same extended gaseous structure. 
Examining Figure~\ref{fig:diagram}, we notice that the absorption system kinematics and metallicities remain similar across the ${\sim} 90\ {h_{70}}^{-1}$~kpc proper ($0.32\ {h_{70}}^{-1}$~Mpc comoving) distance separating the two LOS. The highest column density absorptions in the FG LOS all have $\log\ N$(\ion{H}{I}) $> 18.5$ counterparts in the BG LOS. The main exception is that the $\log\ N$(\ion{H}{I}) $= 20.2$ BG-C component does not correspond to a high $N$(\ion{H}{I}) absorption in the FG LOS. %Cosmic web filaments are expected to consist of clumpy, moderate column density gas distributed over cosmological scales, much like the clouds diagrammed in Figure~\ref{fig:diagram}. The two quasars offer parallel LOS that appear to probe gas clouds distributed along the length of a filament. 

% The abundances are calculated with respect to solar values \citep{2003ApJ...591.1220L}
\begin{table*}
\caption{Abundances relative to solar values \citep[taken from][]{2003ApJ...591.1220L} for the $\log\ N$(\ion{H}{I}) (cm$^{-2}$) $> 18.0$ absorption systems detected along the BG and FG quasar LOS. Zero velocity corresponds to $z = 2.6894$.}
% Abundances for the $\log\  N$(\ion{H}{I}) (cm$^{-2}$) $> 18.0$ absorption systems detected along the two LOS. Velocities are relative to $z = 2.6894$.
\label{table:Met_BG}
\centering
\begin{tabular}{l  r  r  r |r  r  r  r  r  r  r  r  r  r }
\hline \hline
Comp. & $v$ (km $\rm s^{-1}$) & $N$(\ion{H}{I}) & $\sigma_{N(\ion{H}{I})}$ & O & $\sigma_{\rm O}$ & Si & $\sigma_{\rm Si}$ & C & $\sigma_{\rm C}$ & Al & $\sigma_{\rm Al}$ & Fe & $\sigma_{\rm Fe}$\\
\hline
BG-A1 & -1681 & 	19.90 & 0.06	 &       -1.19 & 0.34 & -1.57 & 0.11 & -1.83 & 0.59 & -2.06 & 0.15 & -1.86 & 0.14 \\

BG-A2 & -1431 & 19.67 & 0.26 &       -1.56 & 0.43 & -2.06 & 0.34 & -2.10 & 0.34 & -2.22 & 0.35 & -2.35 & 0.30 \\

BG-B  &  -834 & 	18.41 & 0.23 & $\leq$-1.80 & 0.25 &   --   &   --   &  --  &  -- &   --   &   --   &  --  &  --   \\
BG-C\hspace{5 pt} & 0 &  20.15  &  0.05  & --  &  --  & -0.72 & 0.11 &  --  &  --  & -0.89 & 0.08 & -1.13 & 0.18 \\

\hline \hline

FG-A1 & -1550 & 18.46 & 0.21 & $\leq$-1.65 & 0.24 & -- & -- & -- & -- & -- & -- & -- & --  \\
            
FG-A2 & -1432 & 18.57 & 0.24 & $\leq$-1.76 & 0.27 & -- & -- & -- & -- & -- & -- & -- & --  \\

FG-B  &  -800 & 18.77 & 0.20 & $\leq$-1.96 & 0.23 & -- & -- & -- & -- & -- & -- & -- & --  \\

% $\leq$-0.86 & 0.42
% FG-A2 Limits
% [C/H] = -0.699 +- 1.6252  0.4820
% [Si/H] = -0.478 +- 0.3955 
% [Al/H] = -0.499 +- 0.4470 
% [Fe/H] = -1.080 +- 0.4691 

\hline
\end{tabular}
\end{table*}

In region A, the dense gas extends more than $90\ {h_{70}}^{-1}$~kpc in the transverse direction and 250~km~s$^{-1}$
along the LOS. The components, which have [O/H] $\leq -1.7$ (FG) and [C/O]~$\sim$~[Fe/O]~$\sim$~$-0.5$ (BG), are consistent with very metal poor gas \citep{2014MNRAS.440..307D} and approach what is believed to be the IGM metallicity \citep{2004ApJ...606...92S}.
Each LOS has one strong component in region B.
BG-B and FG-B are closely aligned and also have low metal abundances: [O/H] $\leq -1.8\ \rm and\ -2.0$, respectively.  
Finally, the $\log\ N$(\ion{H}{I}) $\simeq 20.2$ sub-DLA in region C with [Si/H]~$= -0.7$ is likely associated with a galaxy. 
The BG-C abundance is somewhat higher than that of DLAs at the same redshift \citep{2012ApJ...755...89R}. If the BG-C sub-DLA galaxy is accreting gas from its surroundings, this could explain the lack of higher column density absorptions in region C of the FG spectrum. Due to accretion, gas in the galaxy halo becomes more sparsely distributed. 
The A, B, and C regions have distinct properties that are overall consistent in both spectra, but along each LOS the clouds do not appear to be directly in contact.

These absorptions span more than $1\,700\ \rm km\ s^{-1}$ along each LOS, which corresponds to a proper distance of $6.4\ {h_{70}}^{-1}$~Mpc at $z = 2.69$ (23.6~${h_{70}}^{-1}$~Mpc comoving). Velocity differences at this scale are dominated by the Hubble flow, rather than physical velocities intrinsic to the gas clouds.
The A, B, and C absorption regions have a velocity separation of more than $5\,000\ \rm km\ s^{-1}$ from the FG quasar at $z = 2.75$, which makes direct association with the quasar environment unlikely \citep{2010MNRAS.406.1435E}.

In addition to the $\log\ N$(\ion{H}{I})~$> 18.0$ components, several weaker absorptions within the ${\sim} 2\,000\ \rm km\ s^{-1}$ region are common to both LOS. Corresponding absorptions with $\log\ N$(\ion{H}{I})~$ = 14.5 - 15.2$ occur near -1180~km~s$^{-1}$, -720~km~s$^{-1}$, and $-450$~km~s$^{-1}$. For Ly$\alpha$ forest absorptions in the range $\log\ N$(\ion{H}{I})~$ = [14, 17]$ at $z \sim 2.55$, \cite{2013A&A...552A..77K} measured a mean line density $dN/dz = 76.38 \pm 7.32$.
The regions where such Ly$\alpha$ absorptions can be detected in both LOS cover a total of $950\ \rm km\ s^{-1}$. This is less than the full coincident region, since the $\log\ N$(\ion{H}{I})~$> 18.0$ systems completely absorb the flux in the remaining portion of the coincident region.
% the rest of the window is fully absorbed by
% The regions where coincident IGM absorptions can be detected cover a total of ${\sim} 950\ \rm km\ s^{-1}$, where neither the absorption in the FG or BG LOS is at risk of blending with one of the $\log\ N$(\ion{H}{I})~$> 18.0$ systems.
% Such absorptions can be detected in both LOS only when neither the FG nor BG LOS absorption is at risk of blending with one of the $\log\ N$(\ion{H}{I})~$> 18.0$ systems. 
% For a redshift path of 0.012, 
The expected number of low column density absorptions is therefore $0.89 \pm 0.09$, whereas three are observed. The probability of such an occurrence is only 6\%.
% suggest that the diffuse gas between the stronger components is also similarly distributed across the two LOS.

To investigate whether the strong absorption systems imply an overdensity, we evaluate the probability of finding two additional LLS within $2\,000\ \rm km\ s^{-1}$, given that one LLS occurs along the total path length. 
\cite{2013ApJ...765..137O} determined that the line density, $dN/dz$, for log $N$(\ion{H}{I}) $\geq 17.2\ \rm cm^{-2}$ absorptions is $0.92 \pm 0.18$. For the redshift path between the BG quasar at $z = 2.916$ and the end of the spectrum at 3\,000 \AA\ ($z = 1.468$), this probability is ${\sim} 0.07\%$. % that two additional LLS would occur within $2\,000\ \rm km\ s^{-1}$ of a first LLS. 
Since the LLS absorptions in the J0913-0107 spectrum all have log~$N$(\ion{H}{I})~$\geq 18.0$, the ${\sim} 0.07\%$ probability can be considered an upper limit.
The LOS clearly probe an overdense region, which may be evidence of a galaxy protocluster, perhaps with a filamentary structure, or a filament in the IGM. % can be interpreted as
We present arguments for the two interpretations.

Following hierarchical structure formation, regions that give rise to galaxy clusters at $z < 1$ have been matter-rich throughout cosmic time. %  marked by overdensities
In cosmological simulations, individual galaxies come together along gaseous filaments, creating small groups that in turn merge to form clusters by low redshift.
Identifying overdense regions at high redshift that will eventually collapse to form gravitationally bound clusters at $z = 0$ is of particular interest for investigating galaxy cluster evolution. % important, necessary 
By tracking cluster formation in cosmological simulations, \cite{2013ApJ...779..127C} were able to predict the $z = 0$ cluster mass from the galaxy overdensity at $ 2 < z < 5$. 
The comoving length of the coincident absorption region along the J0913-0107 LOS is consistent with the expected effective diameter for a protocluster. However, to be identified as a protocluster at $z \sim 2 - 3$ with 80\% confidence, a (25~Mpc comoving)$^3$ region must exhibit an overdensity of more than twice as many galaxies with $M_{\ast} > 10^9\ M_{\odot}$ than a typical field. 

We consider whether is it likely that the absorbers probe gas in the environment of massive galaxies.
\cite{2014MNRAS.438..529R} associated log~$N$(\ion{H}{I})~$> 17$ absorptions with galaxies in cosmological, hydrodynamical simulations \citep[see also][]{2011ApJ...743...82M} at $z = 3$ and found that most strong absorbers are most closely related to low mass galaxies with $M_{\ast} < 10^8\ M_{\odot}$. Only log~$N$(\ion{H}{I})~$> 21$ absorptions are routinely associated with $M_{\ast} > 10^9\ M_{\odot}$ galaxies. %Although the A, B, and C regions in the J0913-0107 spectra represent an overdensity, 
The mass-metallicity relation similarly suggests that typical DLAs have $M_{\ast} \sim 10^{8.5}\ M_{\odot}$ \citep{2013MNRAS.430.2680M}. Although the A, B, and C regions in the J0913-0107 LOS are overdense, the galaxies may not be sufficiently massive to directly contribute to the protocluster criterion.

Each quasar LOS can potentially detect \ion{C}{IV} gas associated with the circumgalactic medium of massive star-forming galaxies out to a distance of $0.42\ h^{-1}_{70}$~Mpc comoving \citep{2010ApJ...721..174M}.
% at $2.5 < z < 3.0$
Combining the Schechter mass function for field galaxies \citep{2014ApJ...783...85T} with the factor of 2.2 overdensity necessary for a galaxy protocluster \citep{2013ApJ...779..127C}, the LOS probe a volume that would encompass only ${\sim} 0.1\ M_{\ast} > 10^9\ M_{\odot}$ protocluster galaxies if they are randomly distributed.
The possibility that the overdense region intersects a protocluster therefore cannot be ruled out, even if the absorber galaxies are not particularly massive.
However, the overdensity of log~$N$(\ion{H}{I})~$> 18$ absorbers is ${\sim} 90$, which is much higher than the expected overdensity of galaxies in a protocluster. This suggests that the absorbers could be aligned in a filamentary structure.

%The two parallel quasar LOS may probe gas clouds distributed along the length of a filament.
Cosmic web filaments are expected to consist of clumpy, moderate column density gas distributed over cosmological scales (e.g., \citealt{2005MNRAS.359..272C, 2014MNRAS.441.2923C}; see also \citealt{2006A&A...445....1R, 2014arXiv1402.1165D} for examples of gaseous filaments in hydrodynamical simulations), much like the clouds diagrammed in Figure~\ref{fig:diagram}. 
%see also 2006A&A...445....1R, 2014arXiv1402.1165D for examples of gaseous filaments in hydrodynamical simulations.
Furthermore, the process of mass build-up that results in galaxy clusters at low redshift is thought to occur along filamentary structures.
The metallicity distribution along the overdense A, B, and C regions may indicate a filamentary structure. % suggests
Absorptions in regions A and B probe very metal poor gas; their metal abundances are nearly an order of magnitude below that of the BG-C component.
\cite{2013ApJ...770..138L} identify a bimodality in the metallicity distribution of $z \lesssim 1$ LLS and argue that the low-metallicity ($\rm \langle [X/H] \rangle < -1.57 \pm 0.24$) population traces gas accreting along filaments. Similarly, \cite{2013Sci...341...50B} highlight the metallicity difference between a $z = 2.3$ star-forming galaxy and gas detected in a DLA at an impact parameter of 26~kpc. The metallicity and gas kinematics of this system are consistent with a scenario where infalling IGM gas co-rotates in the halo before accreting onto the galaxy disk. 
The gas in the overdense region may be distributed along the length of a filament, probed by the two parallel quasar LOS.

% Additionally, the \ion{C}{IV} content of the gas suggests that the observed substructures are not in interaction.
The observed substructures are likely not in interaction, based on the \ion{C}{IV} content of the gas. In an environment where galaxies are interacting,
we expect \ion{C}{IV} to be conspicuous both because of metal enrichment and higher temperatures. Relatively little \ion{C}{IV} absorption associated with the overdense \ion{H}{I} region is apparent in the spectra (Figures~\ref{fig:BGmet_Fits} and \ref{fig:FGmet_Fits}), and the \ion{C}{IV} absorption associated with the BG-C sub-DLA is relatively weak. Consistent with the metallicity results, the lack of strong \ion{C}{IV} absorption implies that the overdense region is not highly enriched. With high resolution, high S/N spectra, \cite{1996AJ....112..335S} studied the correspondance between \ion{H}{I} and \ion{C}{IV} absorptions. They found that 90\% of $\log\ N$(\ion{H}{I})~$> 15.2$ absorptions have $\log\ N$(\ion{C}{IV})~$> 12.0$. In the range $\log\ N$(\ion{H}{I})~$ = [15, 17]$, the median \ion{C}{IV}/\ion{H}{I} value is $3 \times 10^{-3}$. The detection limit of $\log\ N$(\ion{C}{IV})~$< 12.8$ (BG) and 13.2 (FG) may nevertheless be insufficient to reveal \ion{C}{IV} absorptions associated with the lower column density \ion{H}{I} components.

% Imaging the field to detect star-forming galaxies is essential for determining whether it satisfies the protocluster criterion. 

% It would be very important to image this field to search for galaxies associated with the overdensity and unveil its true nature. 
 
We favor the interpretation that the gas is distributed in a $6.4\ {h_{70}}^{-1}$~Mpc proper filament at $z \simeq 2.69$. The high concentration of gas clouds along the LOS is difficult to explain with the factor of ${\sim} 2$ galaxy overdensity expected in a protocluster and suggests a clumpy, filamentary structure. The metallicities in regions A and B differ by a factor of ten from the metallicity in region C, and the lack of strong \ion{C}{IV} absorption likewise implies that the overdense region is not highly enriched. However, we cannot rule out that this filamentary structure represents the first stages of cluster formation. Imaging this field to search for galaxies associated with the overdensity is essential to unveiling its true nature. 
% The gas may be in the  early stages of large-scale structure build-up. 
Detecting a possible filament in absorption is a step forward in revealing the structure of the IGM on small scales and foreshadows what will be possible with the next generation of 30~m telescopes.
%
%
%
%______________________________________________________________

\begin{acknowledgements}
We sincerely thank Susanna Vergani for her helpful guidance with preparing the observations and reducing the data.
We also thank the anonymous referee for comments that enhanced the paper. % helped to improve the paper. enhanced the discussion. strengthened
\end{acknowledgements}

%
%________________________________________________________________

% for the bibliography, at the end
\bibliographystyle{aa} % style aa.bst
\bibliography{fila_biblio} % your references Yourfile.bib

\end{document}